\documentclass{elsart5p}
\usepackage{graphicx}
\usepackage{amssymb}

\usepackage[sort&compress]{natbib}

\begin{document}

\begin{frontmatter}



\title{A method to measure the resonance transitions between the gravitationally bound quantum states of neutrons in the GRANIT spectrometer}

\author[ILL]{M. Kreuz}
\author[ILL]{V.V. Nesvizhevsky}
\author[ILL]{P. Schmidt-Wellenburg}
\author[ILL]{T. Soldner}
\author[ILL]{M. Thomas} 
\author[pILL]{H.G. B{\"o}rner}
\author[LPSC]{F. Naraghi} 
\author[LPSC]{G. Pignol} 
\author[LPSC]{K.V. Protasov}
\author[LPSC]{D. Rebreyend} 
\author[LPSC]{F. Vezzu}
\author[LMA]{R. Flaminio} 
\author[LMA]{C. Michel} 
\author[LMA]{L. Pinard}
\author[LMA]{A. Remillieux} 
\author[Virginia]{S. Bae{\ss}ler} 
\author[PNPI]{A.M. Gagarski}
\author[PNPI]{L.A. Grigorieva}
\author[Khlopin]{T.M. Kuzmina}
\author[RhodeIsland]{A.E. Meyerovich} 
\author[ISSP]{L.P. Mezhov-Deglin}
\author[PNPI]{G.A. Petrov}
\author[JINR]{A.V. Strelkov}
\author[Lebedev]{A.Yu. Voronin}

\address[ILL]{ILL, 6 rue Jules Horowitz, Grenoble, France, F-38042}
\address[pILL]{previously ILL}

\address[LPSC]{LPSC, 53, rue des Martyrs, Grenoble, France, F-38026}
\address[LMA]{LMA, 22 bd Niels Bohr, Villeurbanne, France, F-69622}
\address[Virginia]{Virginia University, 1101 Millmont Street, Charlottesville, USA, 22904}
\address[PNPI]{PNPI, Orlova Roscha, Gatchina, Leningrad reg., Russia, 188350}
\address[Khlopin]{Khlopin Institute, 28 Vtoroi Murinsky per., St. Peterburg, Russia, 194021}
\address[RhodeIsland]{University of Rhode Island, Kingston, USA, RI-02881}
\address[ISSP]{ISSP, 2 Institutskaia, Chernogolovka, Moscow reg., Russia, 142432 }
\address[JINR]{JINR, 6 Joliot-Curie, Dubna, Moscow reg., Russia, 141980}
\address[Lebedev]{Lebedev Institute, 53 Leninsky pr., Moscow, Russia, 119991 }

\begin{abstract}
We present a method to measure the resonance transitions between the gravitationally bound quantum states of neutrons in the GRANIT spectrometer. 
The purpose of GRANIT is to improve the accuracy of measurement of the quantum states parameters by several orders of magnitude, 
taking advantage of long storage of Ultracold neutrons at specula trajectories. 
The transitions could be excited using a periodic spatial variation of a magnetic field gradient. 
If the frequency of such a perturbation (in the frame of a moving neutron) coincides with a resonance frequency defined by the energy difference of two quantum states, 
the transition probability will sharply increase. 
The GRANIT experiment is motivated by searches for short-range interactions (in particular spin-dependent interactions), 
by studying the interaction of a quantum system with a gravitational field, 
by searches for extensions of the Standard model, 
by the unique possibility to check the equivalence principle for an object in a quantum state 
and by studying various quantum optics phenomena.
\end{abstract}

%\begin{keyword}
%\end{keyword}

\end{frontmatter}

%==========================================================
\section{Introduction}
\label{introduction}
%==========================================================

According to quantum mechanics, a particle bouncing above a perfect mirror in the Earth's gravitational field has discrete energy states. 
This statement is known since the birth of quantum mechanics as the \emph{quantum bouncer} problem. 
Properties of this system are mentioned in many books and pedagogical articles \cite{Breit,Gold,tHaar,LL,Langh,Flug,Gibbs,Lusch1978,Sak,JGea}.
The gravitational quantum states of neutrons in the Earth's gravitational field have been disclosed recently using Ultracold neutrons (UCN)
\cite{Nesv2000,Nesv2002,Bowles2002,Schwarzschild2002,Nesv2003_PRD,Nesv2003_UFN,Nesv2004,Nesv2005_EPJ,Nesv2005_NIST,Nesv2005_review,Voronin2006,Mey2006,Mey2007,West2007}, 
in taking advantage of the unique properties of UCN \cite{Lusch1} and the exceptionally large characteristic spatial size of the wavefunctions of the quantum states $z_0 = \left( \hbar^2 / (2 m^2 g) \right)^{1/3} = 5.9 \ \mu {\rm m}$.
These states bring rich physical information for studying the interaction of a quantum system with the gravitational field 
\cite{Ahl,Khor,Bini,Bertol,Kief,Lecl,AhlK,Ban,Brau,Boul,Buis,Acc,Saha,Man,Sil,Pignol2007,Arminjon2008}. 
For example, they allow to probe the equivalence principle in a quantum regime \cite{Onof1996,Onof1997,Herd,Wawr,Chrys,CentrifugalPRA,CentrifugalNIM}.
Also, due to the smallness of the state energy $\sim 10^{-12}$~eV, 
their observation and study are useful to constrain extra short range interactions \cite{Abele2003,Westphal:2007,NesPr,NesvPrP}. 
It is particularly sensitive to hypothetical axion-like spin-matter interactions \cite{BNPV,BNPPV}.
This phenomenon could also be used to constrain the neutron electric charge \cite{Nesv2005_review}.
Finally, gravitational quantum states of neutrons provide a perfect experimental laboratory for 
surface studies, neutron quantum optics phenomena, such as quantum revivals and localization \cite{Kalb,Rob,Berb,Bell,Math,Witt,Rom,Gonz}.
They may even have a connection with the search for extensions of quantum mechanics \cite{Steyerl1992,Steyerl1996,Steyerl997,Bestle1998,Brenner2000,Nesvizhevsky2005}.

The purpose of GRANIT is to improve the accuracy of measurement of the quantum states parameters by several orders of magnitude, 
taking advantage of long storage of Ultracold neutrons at specula trajectories \cite{Pignol}.
Specific technical developments were undertaken, such as studies for a dedicated UCN source \cite{SchmidtWellenburg3} with a proper extraction system \cite{SchmidtWellenburg1,Barnard2008}, 
and quality tests for neutron mirror reflections \cite{Nesvizhevsky2006,Nesvizhevsky2007}.

We describe a general design of the GRANIT spectrometer in part 2. 
Part 3 presents resonance transitions between the quantum states.
Part 4 is devoted to a description of the first experiment intended to be done with GRANIT: The magnetically-induced resonance transitions between gravitationally bound quantum states of neutrons measured in a flow through mode.

%=====================================================
\section{The GRANIT spectrometer}
\label{GRANIT}
%=====================================================

The GRANIT spectrometer is a new-generation type of gravitational
spectrometer. 
All important parameters are improved with respect to the first version. 
We will 
1) increase considerably the storage time of neutrons in the quantum states \cite{Nesv2005_review,Pignol}; 
2) apply the method of resonance transitions between the gravitationally bound
quantum states of neutrons \cite{Nesv2005_review}; 
3) increase the UCN density using a dedicated $^4$He UCN source \cite{SchmidtWellenburg3}; 
4) profit from a permanent installation of the spectrometer in a more comfortable experimental environment; 
5) use polarized neutrons and polarization analysis. 
All studies previewed for the preceding gravitational spectrometer \cite{Nesv2000}, dismantled at present, would be continued with GRANIT. 
Besides, a new kind of precision experiment would become possible \cite{Nesv2005_review,BNPV}.

The GRANIT spectrometer will be installed at the level C at the ILL reactor (the ground level, close to the reactor core). 
Details of neutron transport, the dedicated UCN source, and neutron extraction are described in \cite{SchmidtWellenburg3}.

\begin{figure*}
\begin{center}
\includegraphics[width=.8\linewidth]{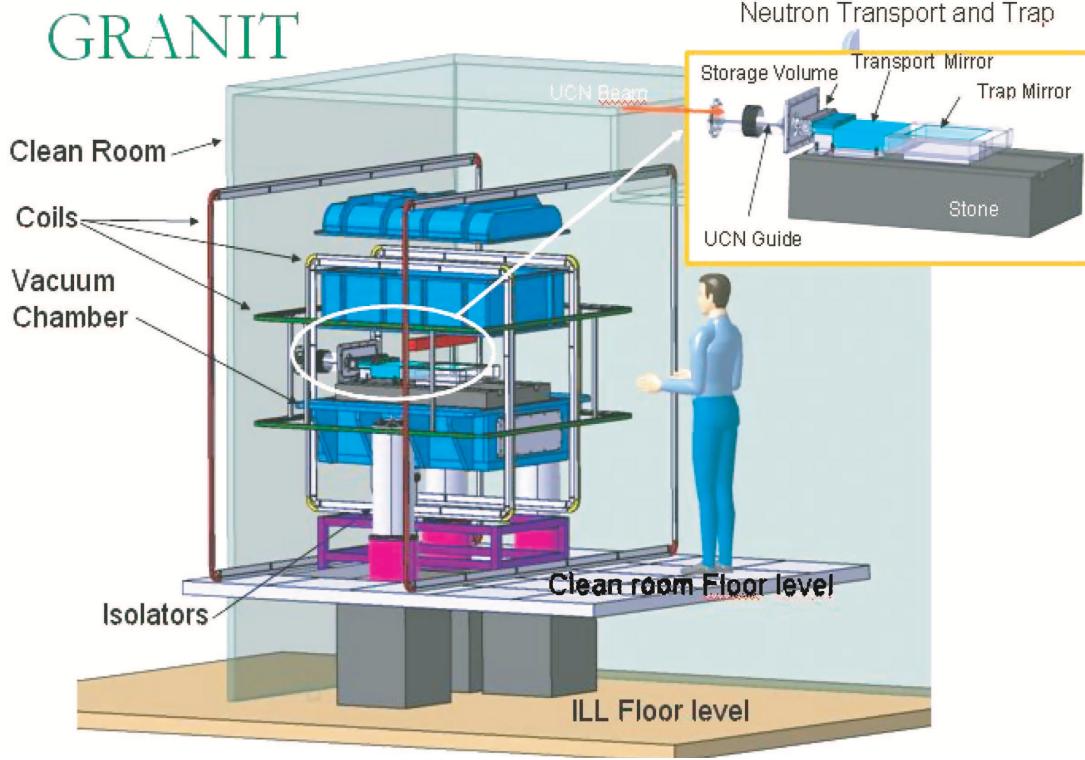}\end{center}

\caption{\label{spectrometer} GRANIT spectrometer.}
\end{figure*}

\begin{figure}
\begin{center}
\includegraphics[width=1.08\linewidth,keepaspectratio]{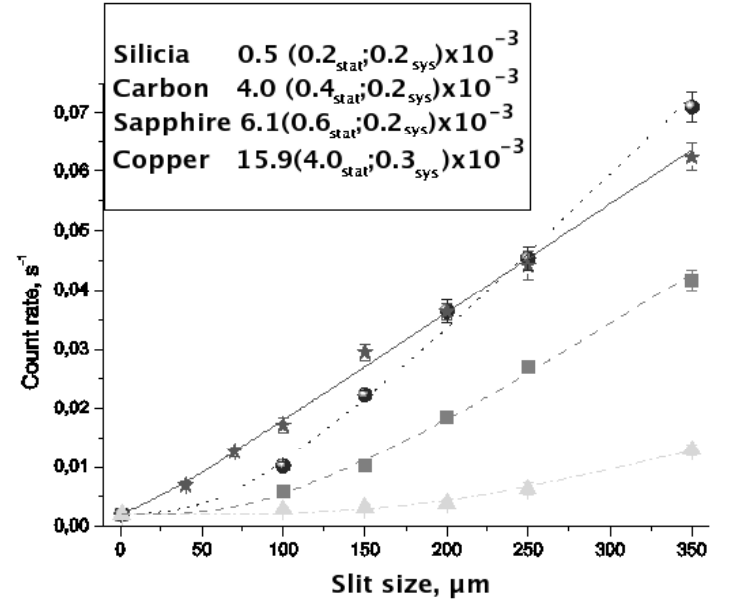}\end{center}

\caption{\label{specular}
Neutron transmission though parallel plates as a function of the distance between the plates. 
The angle at the first reflection was set to 30$^\circ$. 
Deviations from linearity is due to losses, mainly defined by non-specular reflections. 
The measured probability of the total losses per collision is indicated for four tested materials or coatings: 
silica (stars and solid line), 
silica coated with diamond (dots and dotted line), 
sapphire (squares and dashed line), 
and for silica coated with copper (triangles and dashed-dotted line).
}
\end{figure}

The GRANIT spectrometer is shown in Fig. 3 in its simplest configuration.
It consists of the following parts installed in vacuum on a granite table:
\begin{itemize}
\item The neutron transport system connects the UCN source \cite{SchmidtWellenburg3} to GRANIT via an intermediate storage volume. 
Mechanical elasticity of the transport system allows us to decouple GRANIT from the UCN source vibrations.
\item A semi-diffusive narrow slit with a height of 100-200 $\mu$m \cite{SchmidtWellenburg1,Barnard2008} at the exit of the intermediate storage volume allows us to extract UCN with
small vertical velocity components needed for GRANIT. 
Other UCN would be reflected back to the source, thus reducing loss of the UCN density in the source. 
This arrangement allows us to use a $^4$He UCN source in a flow-through mode.
\item A transport mirror is installed at the exit of the semi-diffusive slit.
It consists of a horizontal mirror, two vertical side walls, and a scatterer/absorber above the horizontal mirror. 
The transport mirror selects a few of the lowest quantum states. 
\item A square main mirror with a size of 30 cm and four vertical side
walls is the principle element of the neutron mirror system in GRANIT. 
It should provide large neutron lifetimes in the gravitationally bound
quantum states of neutrons in the storage measuring mode. 
Extremely severe constraints for parameters of this mirror trap are analyzed in \cite{Pignol}. 
Some parameters have already been tested \cite{Nesvizhevsky2006,Nesvizhevsky2007}, motivating the choice of the mirror's material (see fig \ref{specular}). 
\item A resonant transition system for the first experiment in the flow-through
mode with GRANIT consists in a multi-wire assembly installed above the
transport mirror (see part 4). 
It produces a spatially periodic gradient of the magnetic field. 
\item Three types of detectors will be used at the first stage of the GRANIT experiment. 
$^3$He gaseous proportional counters with extremely low background have been developed for this experiment. 
They will be used for integral measurements. 
Position-sensitive nuclear-track UCN detectors with a record resolution of $\sim 1 \ \mu$m \cite{Nesv2000,Nesv2005_EPJ} 
will be used to study the the spatial distribution of neutrons in the quantum states. 
Real-time position-sensitive detectors with a resolution of a few hundred $\mu$m will 
be developed to measure velocity distributions of neutrons in the quantum states.
\item Numerous spectrum-shaping and spectrum-analyzing devices as well as
horizontal leveling and positioning systems will be similar to
those used earlier in 
\cite{Nesv2000,Nesv2002,Bowles2002,Schwarzschild2002,Nesv2003_PRD,Nesv2003_UFN,Nesv2004,Nesv2005_EPJ,Nesv2005_NIST,Nesv2005_review,Voronin2006,Mey2006,Mey2007,West2007}. 
Their parameters, nevertheless, will be significantly improved compared to the preceding versions.
\end{itemize}
~\\
Three main systems will be constructed outside the vacuum chamber to isolate the spectrometer from external perturbations.
\begin{itemize}
\item Anti-vibration and leveling system. 
The first level of protection against vibrations and rough leveling of the installation is provided by
three pneumatic feet with controlled valves, which support the vacuum chamber. 
High level anti-vibration protection and leveling is provided by
three high-load piezo-elements, installed inside the vacuum chamber. 
These piezo-elements support the granite with all optical elements on top of it.
The vibrations associated with ground will be largely suppressed by the system described above. 
The main residual perturbation is expected to come from the mechanical connection between GRANIT and the UCN source. 
This effect has to be studied after installation of the spectrometer. 
One should note that relative position of all optics elements are not affected by vibrations and
slow mechanical drifts anyway, because they all mounted rigidly to the granite table.

\item Clean room. 
The optical elements of GRANIT are very sensitive to dust contaminations. 
First, by precise adjustments with a typical accuracy of $\sim 1 \ \mu$m require dust particles to be smaller than this value. 
Second, by eventual dust particles between two mirror surfaces approaching each
other at such a distance would damage our very delicate and expensive mirrors. 
Therefore the spectrometer is installed inside a clean room. 
The air flow in this room is regulated in such a way that its flow is laminar, without turbulences, in the central zone of the spectrometer, which contains all optical elements.
\item Control of magnetic fields. 
Compensation of local magnetic fields and introduction of a constant field for neutron spin guiding are provided by
three pairs of coils in Helmholtz configuration, installed around the chamber. 
Later, for experiments requiring low values of residual magnetic fields, we will replace the coils by a $\mu$-metal anti-magnetic screen.
\end{itemize}

%==========================================================
\section{Resonant transitions between quantum states}
\label{resonances}
%==========================================================

\begin{figure}
\begin{center}
\includegraphics[width=1.08\linewidth,keepaspectratio]{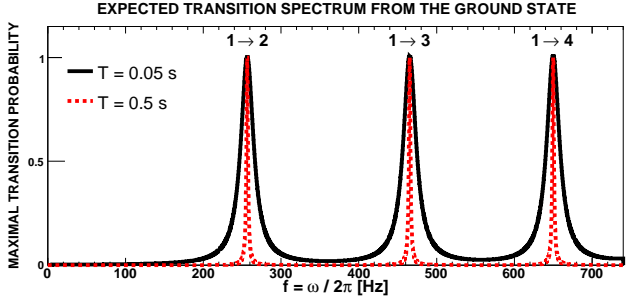}\end{center}
\caption{\label{transitions} Maximum probability for a neutron to leave the ground state is shown as a function of the excitation frequency for two different excitation times.}
\end{figure}

The strategy to measure the spectrum of the gravitational quantum levels is based on resonant excitations of transitions between two states.
Assume that one applies an harmonic perturbation $\hat{V}(t)=Re \left( V(z) e^{i \omega t} \right)$ to a neutron in a pure quantum state $\left|N\right\rangle$.
If the excitation angular frequency $\omega$ is close to that associated with a transition, $\omega_{Nn} = \frac{1}{\hbar} |E_N - E_n|$, 
the probability to excite the neutron to the state $\left|n \right\rangle$ is given by the Rabi formula
\begin{equation}
\label{resonanceformula}
P_{N \rightarrow n}(T) = \frac{\sin^2 \left( \sqrt{(\omega - \omega_{N n})^2 + \Omega_{N n}^2} \ \frac{T}{2} \right)}{1 + \left( \frac{\omega - \omega_{N n}}{\Omega_{N n}} \right)^2 }
\end{equation}
where $\Omega_{N n} = \left\langle n\right| V(z) \left|N\right\rangle$ is the Rabi angular frequency that defines the intensity of excitation for the transition $N \rightarrow n$.
This probability reaches a maximum when the resonance condition $\omega = \omega_{N n}$ is satisfied  (it equals exactly $1$ if the excitation time $T$ is chosen properly). 
This condition allows us to measure the energy differences in the spectrum.
The excitation time $T$ will be given by the experimental conditions, and the excitation amplitude has to be adjusted so that the transition probability at the resonance is unity.
Fig. \ref{transitions} shows the expected resonance curves for a neutron initially in the ground state.
A measurement of the resonance curve gives access to the energy differences in the spectrum, with a resolution limited by 
the resonance width that appears in the denominator of (\ref{resonanceformula}).
This width satisfies the Heisenberg principle; it gives us a rought relation between the eventual resolution $\Delta E$ and the excitation time $T$: $\Delta E \cdot T \simeq h$.
Thus, one aims to maximize the excitation time in order to increase the resolution.
As the excitation time is an essential parameter, we list the characteristic timescales of the problem.
\begin{enumerate}
\item The minimal excitation time needed to resolve the resonances -- the width of each resonance is smaller than the distance between neighboring resonances -- is about 10~ms.
\item A typical UCN -- with an horizontal velocity of about 5~m/s -- 
passes the $30$~cm mirror in $50$~ms.
Thus, we can excite well-resolved transitions in a flow through mode.
This is the purpose of the first measurement in the GRANIT experiment.
\item The neutron $\beta$ decay lifetime is $886$~s, it is the maximal (ultimate) excitation time.
Thus, the spectrum determination is ultimately limited at a relative value of $10^{-6}$ for the first levels.
\end{enumerate}
The next section describes a practical realization of a measurement using resonant transition technique.

%=====================================================
\section{Resonant transitions in the flow through mode}
\label{wires}
%=====================================================

\begin{figure}
\begin{center}
\includegraphics[width=1.08\linewidth,keepaspectratio]{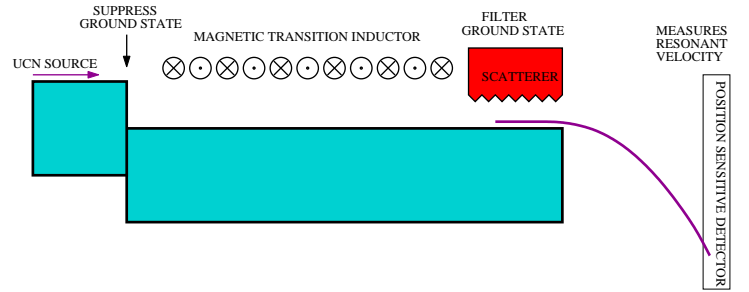}\end{center}

\caption{\label{fluxcontinu} The principle scheme for measuring transitions in a flow through mode.}
\end{figure}

The first measurement with GRANIT will consist in measuring resonant transitions between the gravitational quantum states using a flow through mode, using unpolarized neutrons.
The basic scheme is shown in fig. \ref{fluxcontinu}. 
A neutron will experience different processes in the following order:
\begin{enumerate}
\item A neutron is extracted from the source with a small vertical velocity. Its horizontal velocity is distributed from $\sim 4$~m/s to $\sim 7$~m/s.

\item The neutron passes a negative step with a height of about $h = 20 \ \mu$m. 
The classical vertical energy of the neutron after the step is larger than $m g h$.
It means that the ground quantum state population is largely suppressed.
After the step, the neutron occupies the states $\left|2\right\rangle$, $\left|3\right\rangle$, $\left|4\right\rangle$, etc, though the populations of the ones with lower quantum numbers are also somewhat reduced.
We are going to induce and measure transition from these states to the ground state $\left|1\right\rangle$.

\item The neutron is exposed to a spatially periodic excitation.
This is provided by a magnetic field gradient, induced by horizontal conducting wires.
The current in the wires is set such as to maximize the transition probability at a resonance. 
It is of the order of a few Amperes.
The distance between the wires defines a spatial periodicity of the excitation of $d = 10$~mm.
For the parameters listed above, the magnetic gradient can be considered as a relatively small perturbation.
According to its horizontal velocity $v$, the neutron will experience an effective time-dependent perturbation with a frequency $v/d$.
%The energy exchange to change vertical levels is provided by the horizontal motion, this corresponds to a horizontal velocity change much smaller than the absolute horizontal velocity.
Only the neutrons for which the frequency $v/d$ corresponds to the resonance frequency 
(at the precision of the resonance width) will transite to the ground state.
Thus, the resonance frequency that we aim to measure corresponds to a horizontal velocity.

\item Then the neutron enters a slit between the horizontal mirror and an absorber above. 
This slit filters the ground state; the excited states are scattered away.
 
\item To measure the horizontal resonant velocity, the neutron will fall following an almost classical parabolic trajectory (in fact, the spread of the wave function during free fall has to be taken into account as well).
The fall height is measured using a position sensitive detector.
This measurement allows us to calculate the horizontal velocity value.
\end{enumerate}

The signal to be measured would be a peak in the position sensitive detector.
This sequence will allow us to measure the resonant frequencies of the transitions $3 \rightarrow 1$, $4 \rightarrow 1$, and maybe $2 \rightarrow 1$ and $5 \rightarrow 1$, with an accuracy better than $10 \%$.

%=====================================================
\section{Conclusion}
\label{conclusion}
%=====================================================

We described the spectrometer GRANIT which will be used to measure the resonance transitions between the gravitationally bound quantum states of neutrons, as well as a possible scheme for a first experiment with GRANIT.
The spectrometer is under construction.
It will be installed at the H172 neutron guide at the ILL.
First experiments are expected to start in 2009.

The GRANIT project is being built in framework of ANR (Agence Nationale de la Recherche, France) grant.
The installation of GRANIT at the neutron guide H172 at the ILL is financed by ILL and LPSC, France.
We are grateful to all our colleagues for help and advice.

\bibliographystyle{elsart-num}
\bibliography{granit}

\end{document}